\documentclass[aps,twocolumn,showpacs,superscriptaddress,pra]{revtex4}  
\usepackage{graphicx, subfigure}  
\usepackage{dcolumn}   
\usepackage{bm}        
\usepackage{braket}
\usepackage{amssymb}   
\usepackage{enumerate}
\usepackage{amsmath}
\usepackage{float}

\begin{document}

\title{Realizing Topological Transition in a Non-Hermitian Quantum Walk with Circuit QED}
\author{Yizhou Huang}
\author{Zhang-qi Yin} \email{yinzhangqi@mail.tsinghua.edu.cn}
\affiliation{Center for Quantum Information, Institute for Interdisciplinary Information Sciences, Tsinghua University, Beijing 100084, China}
\author{W. L. Yang}
\affiliation{State Key Laboratory of Magnetic Resonance and Atomic and Molecular Physics, Wuhan Institute of Physics and Mathematics, Chinese Academy of Sciences, Wuhan 430071, China}

\begin{abstract}
We extend the non-Hermitian one-dimension quantum walk model [PRL 102, 065703 (2009)] by taking dephasing effect into account.
We prove that the feature of topological transition does not change even when dephasing between the sites within units is present. The potential experimental observation of our theoretical results in the circuit QED system consisting of superconducting qubit coupled to a superconducting resonator mode is discussed and numerically simulated. The results clearly show a topological transition in quantum walk, and display the robustness of such a system to the decay and dephasing of qubits. We also discuss how to extend this model to higher dimension in the circuit QED system.

\end{abstract}

\maketitle

\section{Introduction}

As a quantum analog of the well-known classical random walk, quantum walk serves as a fascinating framework for various quantum information processes, such as basic search \cite{QSM}, universal quantum computation \cite{UCP}, quantum measurement \cite{Xue15} etc. Apart from its numerous applications, quantum walk also displays new traits different from its classical counterpart, such as the fast spreading of the wave function compared to a classical random walk \cite{ACM1}, which was used for explaining the high efficiency photosynthetic energy transfer assisted by
environment \cite{Engel2007,Mohseni,Ai13,Ai14}.
Ref. \cite{RL1} considered an one-dimensional (1D) quantum walk on a bipartite lattice, where a topological transition was discovered
 and experimental implementation in quantum dots or cavity QED (CQED) was briefly discussed. In it, the environment effect was included by using a non-Hermitian Hamiltonian, as done in \cite{Bender07}.
Later, the same model was extended to multi-dimensional systems \cite{RL2}.

\begin{figure}[tbp]
\centering
\includegraphics[width=0.4\textwidth]{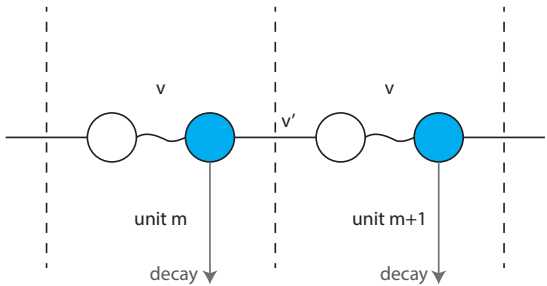}
\caption{(color online) Non-Hermitian quantum walk model. A particle can hop between the nearest sites with strength $v$ or $v^{\prime }$ depending on whether that hop crosses unit boundaries.
When the particle is on the blue sites, it would decay with rate $\protect\gamma $.}
\label{figTheory}
\end{figure}
As shown in Fig. \ref{figTheory}, the quantum walk could be realized on an 1D bipartite indexed lattice where decay sites (blue) and non-decay sites (white) appear in turn. The strengths of hopping between sites are characterized by $v$ (within the same unit) and $v^{\prime }$ (between neighboring units). Due to it, a random \textquotedblleft walker\textquotedblright\ starting from a non-decay site of unit $m_{0}$ may end up decaying from the system from another unit $m$ with probability $P_{m}$. Given the relative strengths of $v$ and $v^{\prime }$, the average displacement of a \textquotedblleft walker\textquotedblright\
\begin{equation}
\sum_{m}m\cdot P_{m}-m_{0}
\end{equation} 
before it decays is quantized as an
integer ($0$ for $v^{\prime }<v$ and $-1$ for $v^{\prime }>v$) no matter
what the decay co-efficient $\gamma >0$ is.
It's worthwhile to note that the average displacement would be \begin{equation}-\frac{v'}{v'+v} \label{eq:classical}\end{equation}
if the system is classical and hoppings are incoherent. One can derive this from the symmetry of the system. \cite{FOOTNOTE3}

As shown in Ref. \cite{LCPT}, such system displayed Parity-Time Symmetry \cite{PT,PT1,PT2}. These
different displacements correspond to unbroken and broken Parity-Time Symmetry regimes and display a topological transition.

In this work we will show that such transition is robust even in the presence of qubit dephase.
Note that this idea can be experimentally realized in a CQED setting, where a qubit is coupled to one resonator mode \cite{Walther06,Yin07}. Specifically, the two states of the qubit ($\ket{e},\ket{g}$) represent two sites of the same unit, where the higher-energy qubit state ($\ket{e}$) has decay rate $
\gamma $.
The cavity photon number state $|n\rangle $ represents the $n$-th unit. Here we suppose that the cavity decay $\kappa $ is much less than qubit decay $\gamma$, therefore we neglect it. Such a system could be an atom in a cavity coupled to one cavity mode. Or similarly as CQED systems: a superconducting Josephson junction which acts as a qubit coupled to a lumped LC oscillator in a superconducting electric circuit. Recently, circuit QED system attracts a lot of attention, as it could be easily scaled up and controlled \cite{QIPC}. We will stick to the circuit QED system setup from now on.

On the other hand, there was a experiment realizing bipartite non-Hermitian quantum walk in optical waveguide \cite{ZREX}, which in essence is a classical simulation of a quantum effect. However, the non-Hermitian quantum walk in circuit QED is fully quantum. Another merit of circuit QED system is that extending to higher dimension quantum walk is relatively easy, as a high-dimensional quantum walk can be implemented by coupling the qubit to more than one resonator mode.  We will show that novel features of a quantum walk emerges with the growth of dimensionality.

In Sec.II, we explained in detail the system setup, and the properties of the effective Hamiltonian of the system. In Sec.III, we extend the analytical theory in \cite{RL1} to accommodate the existence of dephase, and show that the qubit dephase, one major source of decoherence in normal circuit QED systems, won't affect the topological structure experimentally observed. In Sec.IV, we carried out numerical calculations for different parameters of the system, including the preparation of the system, the decay rate of the qubit, the detuning of the system and the effect of qubit dephasing. In Sec.V, we extended this model to higher dimensions of bipartite quantum walk, and carried out numerical calculations for 2D quantum walk specifically. We will show that such quantum walk realization is feasible in circuit QED setting, and good topological transition can be observed with modest requirement on the system parameters afore-mentioned.

\section{The Circuit QED Model}
We consider a system where a two-level qubit is coupled to a resonator mode, with external microwave driving. Such a system can be expressed in standard circuit QED Hamiltonian
\begin{equation} H_{\text{org}} = g \left(\sigma^+ \tilde{a} + \sigma^- \tilde{a} ^\dagger\right) + \Omega/2(\sigma^+ + \sigma^-) + \frac{1}{2} \Delta \epsilon \sigma^z ,
\end{equation}
where $\sigma^z, \sigma^{\pm}$ are Pauli operators for the qubit, $g$ characterizes the coupling strength, and $\tilde{a}(\tilde{a}^\dagger)$ are lowering (rising) operators on the resonator mode in a rotating frame, $\Delta \epsilon$ is (real) detuning of the system, consists of the energy differences between the two states of the qubit minus the minimum energy gap for the resonator, $\Omega$ is the rabi frequency of external drive on the atom.
We put subscript ``org'' to distinguish this Hamiltonian from the effective Hamiltonian $H$ that we will put forward later on.

To consider the dissipation of the system, we only consider qubit decay and dephase.
We use $\gamma$ to characterize qubit decay rate, and $d$ for qubit dephase rate. 
The master equation of the system is
\begin{equation}\label{eq:master_org}
\begin{aligned}
\dot{\rho}(t) = & -i [H_{\text{org}},\rho(t)] +d \left(\sigma^z \rho(t) \sigma^z - \rho(t)\right)
\\ & + \frac{\gamma}{2}\left(2\sigma^- \rho(t) \sigma^+  -\sigma^+ \sigma^- \rho(t) -  \rho(t)\sigma^+ \sigma^-\right),
\end{aligned}
\end{equation}

In usual experimental setup \cite{EX1}, the strength of cavity decay is at the magnitude of $\sim 1$ kHz or less, which is significantly smaller than other parameters in the system, such as the coupling strength $g$ which could reach $\sim 100M$Hz and qubit decay $\gamma$ is in the order of MHz. Therefore, we ignore the cavity decay effects. Besides, the external drive $\Omega$ is easily tunable to fulfill the topological transition conditions.

Experimentally, one can detect qubit decay by either a photon detector or ways latter discussed in this section, but generally one cannot detect qubit dephasing from leaking photons. Furthermore, the setup of this experiment requires immediate measurement of cavity photon number upon qubit decay, which could be realized by detecting photon emission with photon detectors. Therefore, we treated qubit decay and dephase differently. We follow the quantum jump approach in \cite{Plenio98} and put the qubit decay term into the Hamiltonian, which yields an effective non-Hermitian Hamiltonian
\begin{equation} H = g \left(\sigma^+ \tilde{a} + \sigma^- \tilde{a} ^\dagger\right) + \Omega/2(\sigma^+ + \sigma^-) + \frac{1}{2} \Delta \epsilon \sigma^z -  \frac{i\gamma}{2} \ket{e} \bra{e}.
\end{equation}
The details of this derivation are in the appendix. Then Lindblad master equation of the system would be \cite{QM}
\begin{equation}\label{eq:master}
\begin{aligned}
\dot{\rho}(t) =& -i [H,\rho(t)] +d \left(\sigma^z \rho(t) \sigma^z - \rho(t)\right).
\end{aligned}
\end{equation}

Ideally, if the system is initialized at photon number state $|N\rangle$ (which means the ``walker'' starts at exact site $N$), and stays close to $N$ during the whole evolution process, then we can use the approximation
\begin{equation}\label{eq:approx} \tilde{a}\approx \sum\sqrt{N}\ket{n-1}\bra{n}, \tilde{a}^\dagger  \approx \sum \sqrt{N}\ket{n}\bra{n-1}\end{equation} to simplify the operators. The conditions on which this approximation holds will be discussed later.
Then, assuming $\Delta \epsilon \approx 0$, we have \begin{equation} v = \Omega/2, v' = g \sqrt{N}.\end{equation}
Thus, when $v'<v$ ($\frac{\Omega/2}{g} > \sqrt{N}$) the average photon number upon measurement should be $N$ (no change), and when $v' < v$ ($\frac{\Omega/2}{g} < \sqrt{N}$) the average photon number upon measurement should be $N-1$ (change of $-1$).

Due to the existence of decay term $- \frac{i\gamma}{2} \ket{e} \bra{e}$ 
the Hamiltonian is not Hermitian. However, it satisfies the condition of Parity-Time symmetry, that is, after undergoing parity ($\hat{x} \rightarrow -\hat{x}, \hat{p} \rightarrow -\hat{p}$) and time reversal ($\hat{p} \rightarrow -\hat{p}, i \rightarrow -i$) transformation, the new Hamiltonian $H'$ is only a constant away from the old Hamiltonian ($H' = H + c I$). And recent work \cite{LCPT} has proven that when we have $\frac{\Omega/2}{g} > \sqrt{N}$ the eigenvalues of the Hamiltonian are totally real, while $\frac{\Omega/2}{g} < \sqrt{N}$ there's a pair of complex energy eigenvalues, which correspond to unbroken and broken PT symmetries respectively.

As a key element of this experimental realization, a measurement of the photon number in the cavity needs to be carried out immediately after the qubit decays for the first time. One way to implement such scheme is to prepare a low-decay qubit that's coupled to another resonator mode with big decay. By continuous measurement of that other resonator mode \cite{Sun}, which provides much of the qubit decay, one can observe such a decay immediately after it takes place, and thus carry out an energy measurement in the main resonator mode to retrieve photon number. 

\section{Analytical Theory}
The analytical theory for this topological transition has been explained in detail before \cite{RL1}. Here, we will extend their analytical theory to include the qubit dephase.

If the system is pure, one could use $\psi_n^g$ to denote the amplitude at $\ket{g}\otimes \ket{n}$ and $\psi_n^e$ at $\ket{e}\otimes \ket{n}$.
 In the presence of dephase the density matrix of the system is mixed.
We can similarly use density matrix as
$\rho_{n_1,n_2}^{g,g}(0) = \rho_{n_1,n_2}^{e,g} (0)= \rho_{n_1,n_2}^{g,e}(0) = 0$, $\rho_{n_1,n_2}^{e,e}(0) = \delta_{n_1,0}\delta_{n_2,0}$,
where for $c,c'\in\{g,e\}$, $\rho_{n_1,n_2}^{c,c'}$ is the matrix entry corresponding to $\ket{c}\bra{c'} \otimes \ket{n_1}\bra{n_2}$.
We have
\begin{equation} \langle \Delta n \rangle = \sum_n n \cdot \left(\int_0^\infty \gamma \rho_{n,n}^{e,e}(t) dt\right). \end{equation}

Then we translate this to momentum space by having
\begin{equation}
\rho_{n,n'}^{c,c'} = \frac{1}{(2\pi)^2} \oint dk \oint dk' e^{ikn}e^{-ik'n'} \rho_{k,k'}^{c,c'}.
\end{equation}
This Hamiltonian becomes separable as for a $2\times 2$ subspace of $\ket{g/n}\otimes\ket{k}$ we have \begin{equation} H_k = \left(
\begin{array}{cc}
 0 & v_k \\
 v_k^* & \Delta \epsilon - i \gamma / 2 \\ 
\end{array}
\right)
\end{equation} with $v_k = \frac{\Omega}{2} + g\sqrt{N}e^{-i k}$.

Using integration by part, we have that
\begin{equation} n \rho_{n,n}^{e,e} = \frac{i}{(2\pi)^2} \oint dk \oint dk' e^{i(k-k')n} \partial_1\rho_{k,k'}^{e,e}.\end{equation}
Summing over $n$ and integrate over time, we have
\begin{equation} \langle  \Delta n \rangle = i \gamma \int_0^\infty dt \oint \frac{dk}{2\pi} \partial_1 \rho_{k,k}^{e,e}, \label{BASICE1}\end{equation}
where we use $\partial_1 \rho_{k,k}^{e,e}$ as a shorthand of $\frac{\partial}{\partial k'} \rho_{k',k}^{e,e} |_{k'=k}$.

Now, one can define $p_k(t):=\rho_{k,k}^{g,g} + \rho_{k,k}^{e,e}$ as the probability that the subsystem of momentum $k$ hasn't decay till time $t$, with $\partial_t p_k = -\gamma \rho_{k,k}^{e,e}$. Also, one can use polar decomposition as $\rho_{k,k'}^{e,e}(t) = u_{k,k'}(t)\cdot e^{i\theta_{k,k'}(t)}$. With that, one can write
\begin{align}
 \langle  \Delta n \rangle = \frac{i \gamma}{2\pi} \int_0^\infty dt \oint dk & \left(  e^{i\theta_{k,k}(t)} \partial_1 u_{k.k}(t) \right. \notag \\ &  \left. + u_{k,k}(t)\cdot i \cdot e^{i\theta_{k,k}(t)} \cdot \partial_1 \theta_{k,k}(t)  \right) .
 \end{align}
 Considering the fact that $\theta_{k,k}(t) = 0$ as the diagonal terms of a density matrix are real, and $\partial_1 u_{k,k}(t) = \partial_2 u_{k,k}(t)$, one can find that the first term $e^{i\theta_{k,k}(t)} \partial_1 u_{k.k}(t)$ induces an integration of a closed contour, and is thus zero. We can reach
 \begin{equation}
 \langle  \Delta n \rangle = \frac{1}{2\pi} \int_0^\infty dt \oint dk\left( \partial_t p_k(t) \cdot  \partial_1 \theta_{k,k}(t)  \right) .
 \end{equation}

 Defining \begin{equation} I_0 = \oint \frac{dk}{2\pi} \left(p_k \partial_1 \theta_{k,k}(t)|_0^\infty \right),\end{equation} we have through integration by part
 \begin{equation}\langle \Delta n \rangle = I_0 - \int_0^\infty \oint \frac{dk}{2\pi} p_k \partial_t \partial_1 \theta_{k,k}(t).\end{equation}
Given the way we conduct Fourier Transfer, we have $\rho_{k,k'} = \rho_{-k',-k}$, which means $\theta_{k,k'} = \theta_{-k',-k}$. Thus we have \begin{align} \oint \frac{dk}{2\pi} p_k \partial_t \partial_1 \theta_{k,k}(t) = & - \oint \frac{dk}{2\pi} p_k \partial_t \partial_2 \theta_{-k,-k}(t) \notag \\= &  \oint \frac{dk}{2\pi} p_k \partial_t \partial_2 \theta_{k,k}(t). \end{align}
Here we also use the fact that $p_k$ is even function in $k$. Given the fact that $\partial_1 \theta_{k,k}(t) + \partial_2 \theta_{k,k}(t) = 0$, that integration yields zero, and we have
$\langle \Delta n \rangle = I_0$.

Given that the system eventually decays completely, $p_k(t\rightarrow \infty) = 0$, we have
\begin{equation} \langle \Delta n \rangle = - \oint \frac{dk}{2\pi} \left.\frac{\partial \theta_{k',k}(0)}{\partial k'} \right|_{k' = k}.\end{equation}

Since $\rho(0)$ is pure and diagonal in the basis of $\sigma^z$, for $\rho(\epsilon)$ consider up to the first order of $\epsilon \rightarrow 0^+$, one could find that the dephase term $d \left(\sigma^z \rho(t) \sigma^z - \rho(t)\right)$ doesn't affect $\rho(\epsilon)$ to the first order. Thus, by treating $\rho(\epsilon)$ as a pure state (which is similar to the case in Ref. \cite{RL1}), we have
\begin{equation} \langle \Delta n \rangle =  -\oint \frac{dk}{2\pi }\frac{\partial \text{arg}(-i v_k^*)}{\partial k}.\end{equation} The topological structure is that,  if $\Omega/2>g\sqrt{N}$, then the integration of $-i v_k^*$ doesn't contain the axis origin and $\langle \Delta n \rangle = 0$;  if $\Omega/2 < g\sqrt{N}$, then the integration of $-i v_k^*$ is an anti-clockwise contour of a circle centered at $-i \Omega/2$ and with radius $g\sqrt{N}$, which contains the axis origin, thus $\langle \Delta n \rangle = -1$.

\section{Numerical Results and Discussions}
We use a simple numerical integration technique to do the simulation. A $\text{MAXN} = 320$\cite{FOOTNOTE1} (which is the total dimension of the Hilbert space under consideration) dimensional complex vector $V$ (matrix, if dephase is in consideration) is used to store the state of the system. In the absence of dephase, a matrix $U = e^{- i \cdot H \cdot t}$ is computed and $U\cdot V$ simply yields the new state vector $V$. In the presence of qubit dephase we use Lindblad master equation (\ref{eq:master}). Integration is carried out along the way until the amplitude of $V$ converges \cite{FOOTNOTE2}. The time interval $t$ shrinks by half each time and Richardson Extrapolation is carried out until the final result converges.

Suppose one finds that when the qubit decays, the photon number in the resonator is $n$ with probability $P_n$, we define \begin{equation}\langle N \rangle  = \sum_{n=0}^\infty n P_n\end{equation} to be the average photon number upon qubit decay.

\begin{figure}[hbtp]
	\centering
	\includegraphics[width=0.4\textwidth]{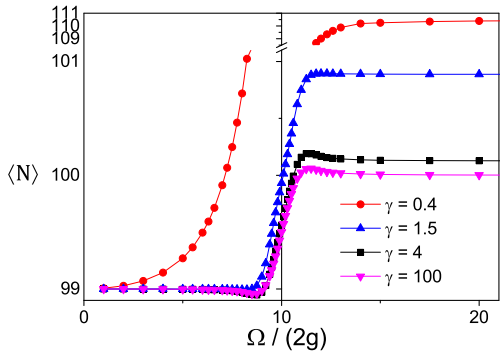}
	\caption{(color online) Results of a non-Hermitian quantum walk for N=100 with different starting distributions
 (Poisson / Fock) of the resonator, and comparison with classical incoherent hopping. The variable $\Omega / 2g$ captures the relative strengths of inter and intra unit hopping in the quantum walk, and $\langle N \rangle$ is the average photon number upon qubit decay. The other parameters are $g = 1$ (set as a benchmark), $\gamma = 4, \Delta \epsilon  = 10^{-4}$ while $\Omega$ varies. For classical incoherent hoppings, we used the result of Eq.(\ref{eq:classical}).}
	\label{figN100}
\end{figure}

A first glimpse of the results are provided in Fig. \ref{figN100} with $N=100$ (which is large, as proposed in \cite{RL1} to meet the assumption of (\ref{eq:approx})).
Clear topological transition is observed if the system is initialized in a Fock state (black) ($|N\rangle$ in number basis).

Since Fock states with $N>>1$ are usually hard to prepare, we considered an alternative : coherent state (Poisson distribution $\sum_k \frac{N^k e^{-N}}{k!}|k\rangle$ in number basis).
However, the simulation (red line in Fig. \ref{figN100}) turns out to be quite similar in the case of non-coherent hopping (classical walker) despite large initial $N$. This shows that classical driving induced coherent states cannot be used for this topological transition.

\begin{figure}[hbtp]
	\centering
	\subfigure[\label{figN1}]{\includegraphics[width=0.23\textwidth]{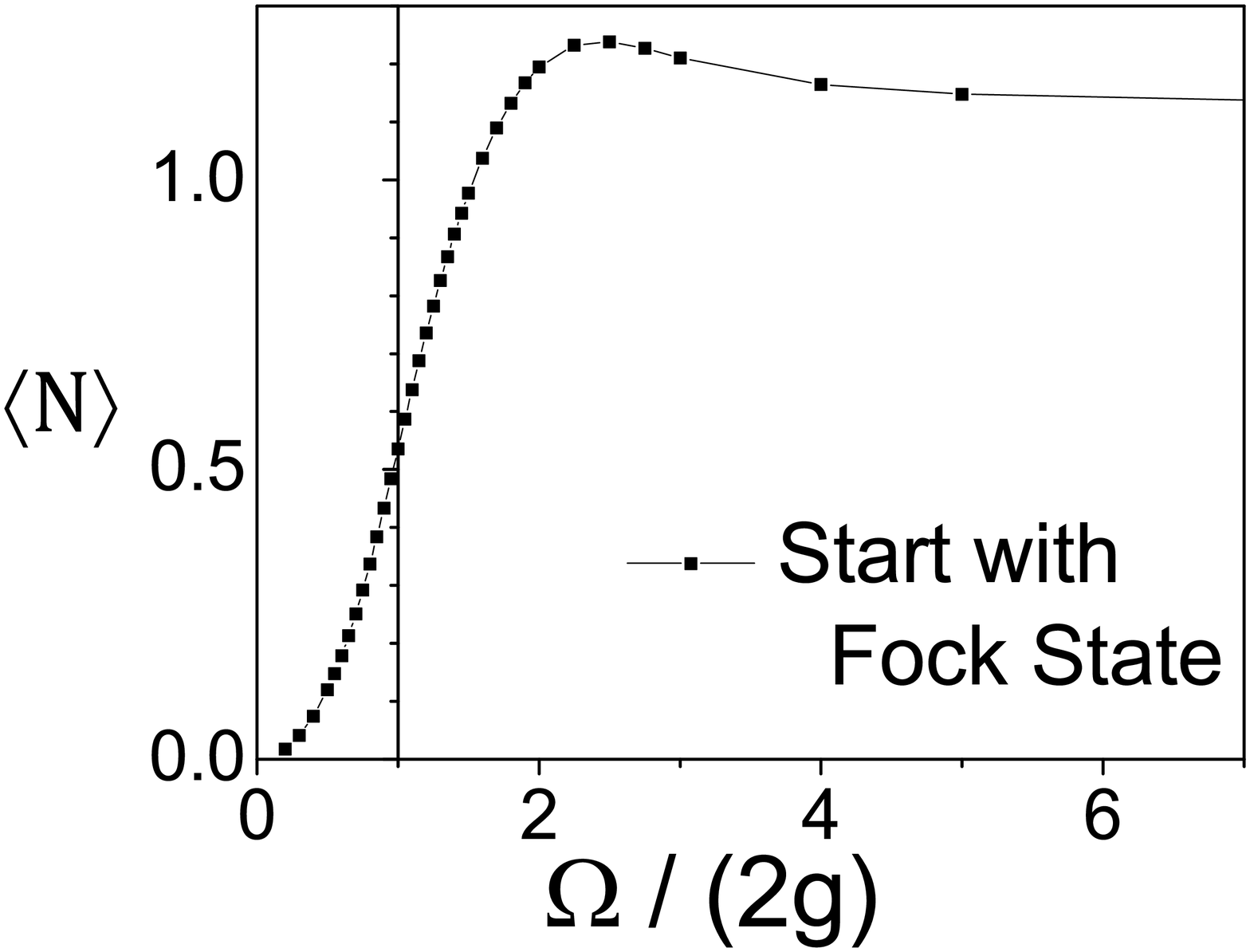}}
	\subfigure[\label{figOverlap}]{\includegraphics[width=0.23\textwidth]{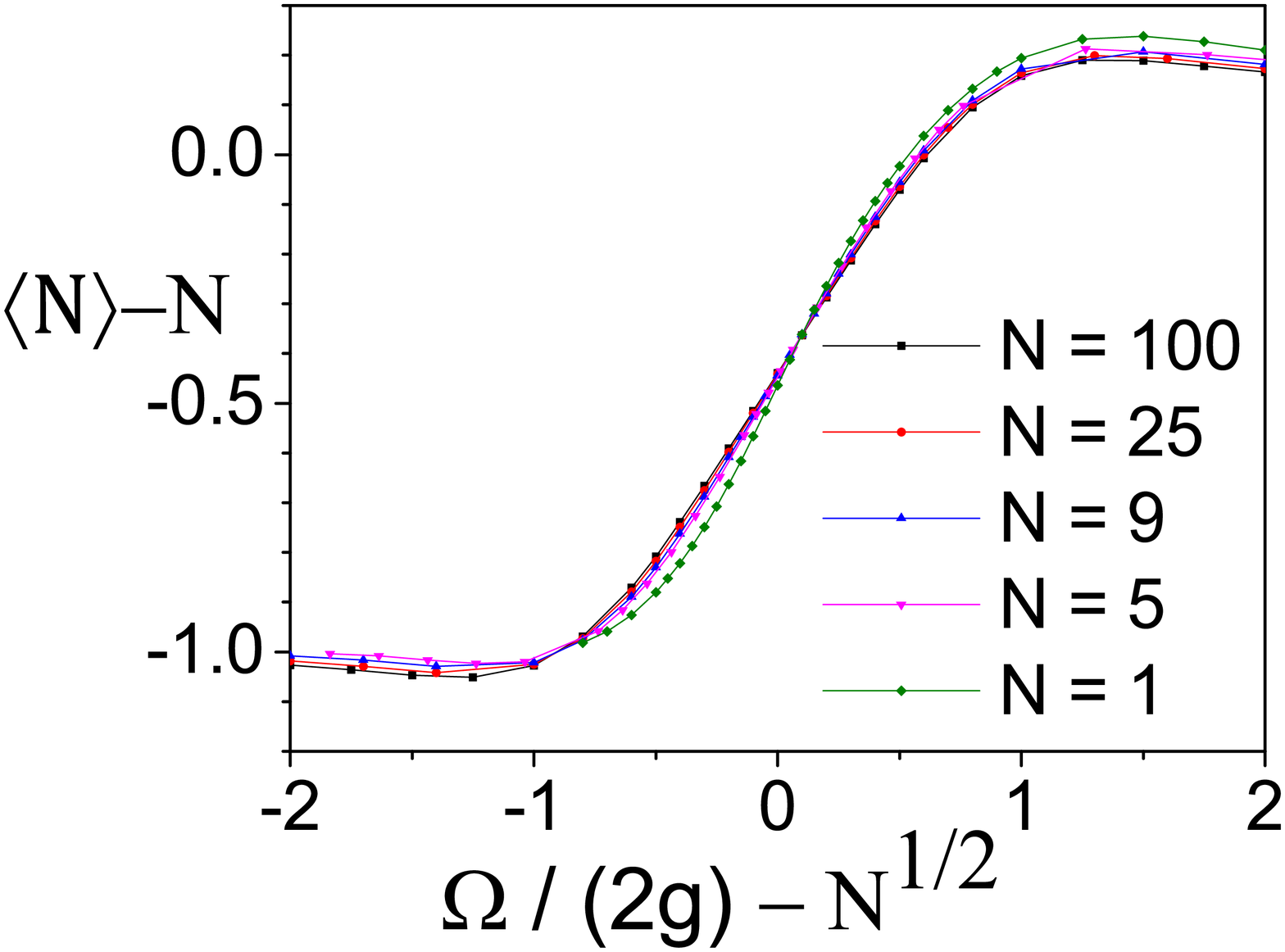}}
	\caption{(color online) Results of simulation for non-Hermitian quantum walk with small initial $N$'s and comparison.
In (a), the variable $\Omega / 2g$ captures the relative strengths of inter and intra unit hopping in the quantum walk, and $\langle N \rangle$ is the average photon number upon qubit decay. In (b), those two variables are compared against topological transition threshold $\sqrt{N}$ and initial photon number $N$ respectively.
The parameters in both figures are (same as Fig. \ref{figN100}) $g = 1, \gamma = 4, \Delta \epsilon  = 10^{-4}$ while $\Omega$ varies.
}
\end{figure}

Considering the fact that Fock states with $N>>1$ are hard to come by, we relax that condition and examine some cases with small Fock numbers. An extreme case of $N = 1$ is displayed in Fig. \ref{figN1} which still preserves the basic traits of a topological transition despite the small initial Fock number.
Further investigation into the problem yields Fig. \ref{figOverlap}, which shows that the curve for different $N$'s overlap.

\subsection{Different Decay Factors and Energy Level Differences}
In this section we consider the effect of qubit decay factors $\gamma$ and detuning $\Delta \epsilon$ on this experiment.

\begin{figure}[hbtp]
	\centering
	\subfigure[]{\includegraphics[width=0.4\textwidth]{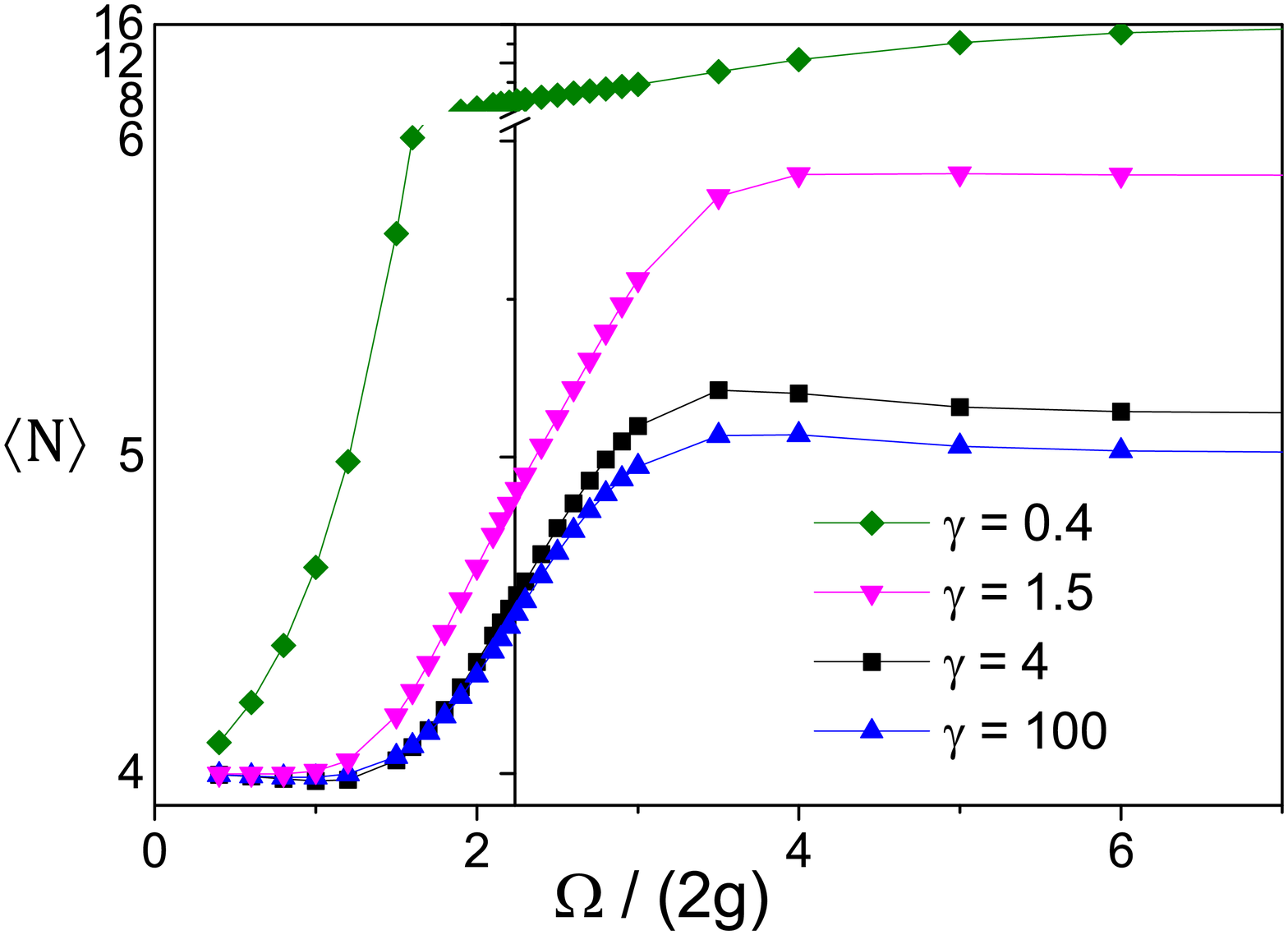}}
	\subfigure[]{\includegraphics[width=0.4\textwidth]{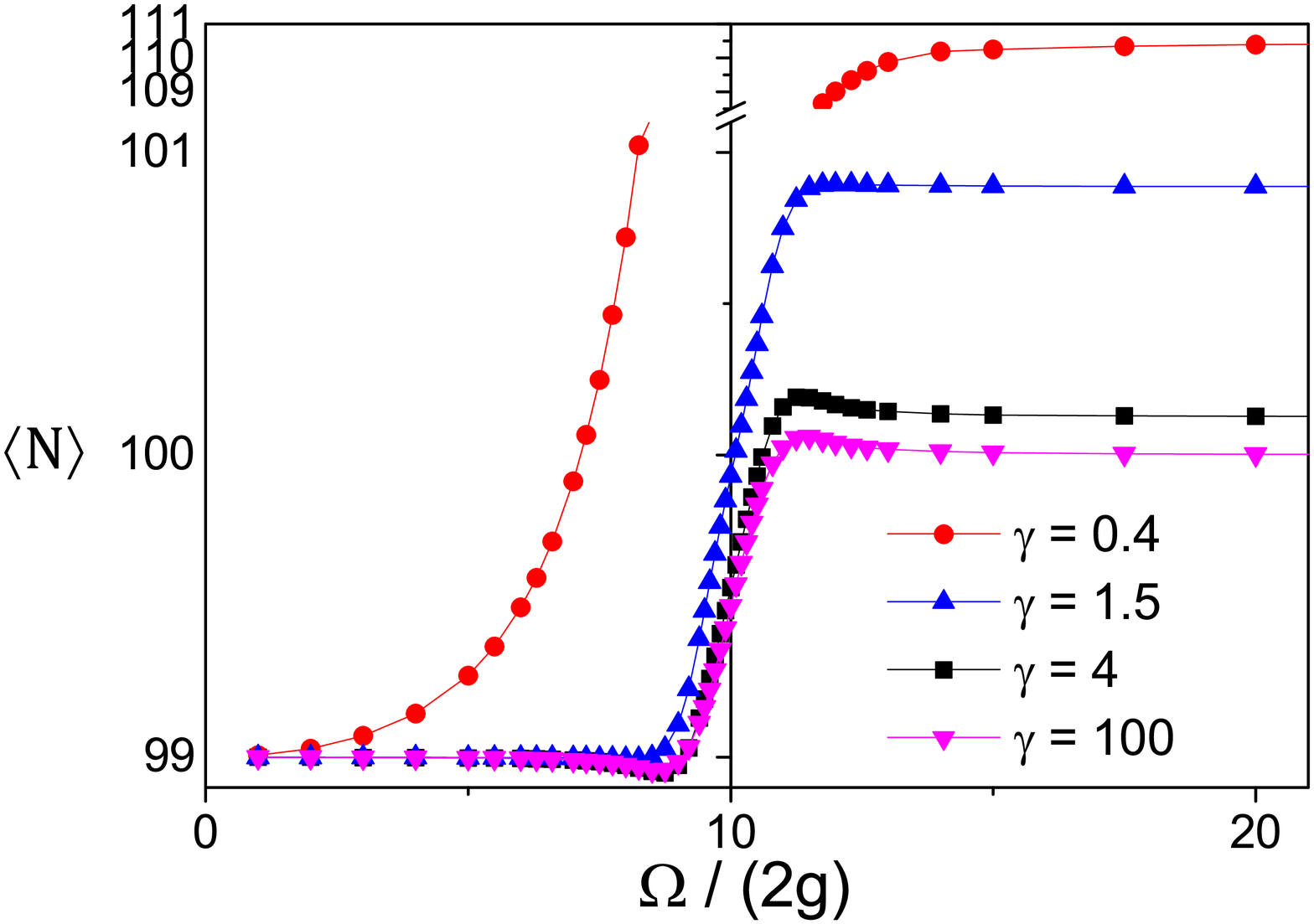}}
	\caption{(color online) Results of simulation for non-Hermitian quantum walk with different qubit decay factor $\gamma$'s under (a) initial $N=5$ and (b) initial $N = 100$. The  variable $\Omega / 2g$ captures the relative strengths of inter and intra unit hopping in the quantum walk, and $\langle N \rangle$ is the average photon number upon qubit decay. The other parameters are $g = 1,  \Delta \epsilon =10^{-4}$ while $\Omega$ varies. One can observe that if qubit decay factor $\gamma$ is small, the curve significantly deviates from theoretical deduction, while for large $\gamma$'s, the curves converge.}
	\label{figGamma}
\end{figure}

In theory \cite{RL1}, qubit decay factor $\gamma$ only affects the (expected) evolution time, not the topological transition, so the expected result should be the same for different $\gamma$'s. We run a simulation for different initial $N$'s with identical parameters like $g$ and $\Delta \epsilon$ for different decay factors $\gamma$'s, with result in Fig. \ref{figGamma}.
We observe that the qubit decay factor $\gamma$ cannot be too small (at least $\gamma >  g$ should be satisfied), otherwise the photon number upon measurement would be too large. This is necessary for both small and large $N$'s.

The difference between analytical theory and numerical results lies in the assumption of equation (\ref{eq:approx}), which is only valid if the system stays close to photon number $N$ throughout the whole evolution. If decay factor $\gamma$ is small, the quantum walker can walk far away from the initial site, overturning that assumption.
Specifically, since we have \begin{equation}
\tilde{a}= \sum_n \sqrt{n}\ket{n-1}\bra{n}, \tilde{a}^\dagger = \sum_n \sqrt{n}\ket{n}\bra{n-1},\end{equation} the strengths of inter-site hoppings grow stronger as the
site number (or energy level in our simulation) 
increases, which explains why for small $\gamma$'s the curves deviate significantly to higher energy levels.

Luckily, big $\gamma$'s does not present an experimental difficulty in reality, as big qubit decays are usually easy to generate. We found that $\gamma = 4$ (4 times the coupling strength) is roughly where qubit decay rate starts to negatively affect the observed topological transition, we will stick to this value from now on.

\begin{figure}[hbtp]
	\centering
	\includegraphics[width=0.37\textwidth]{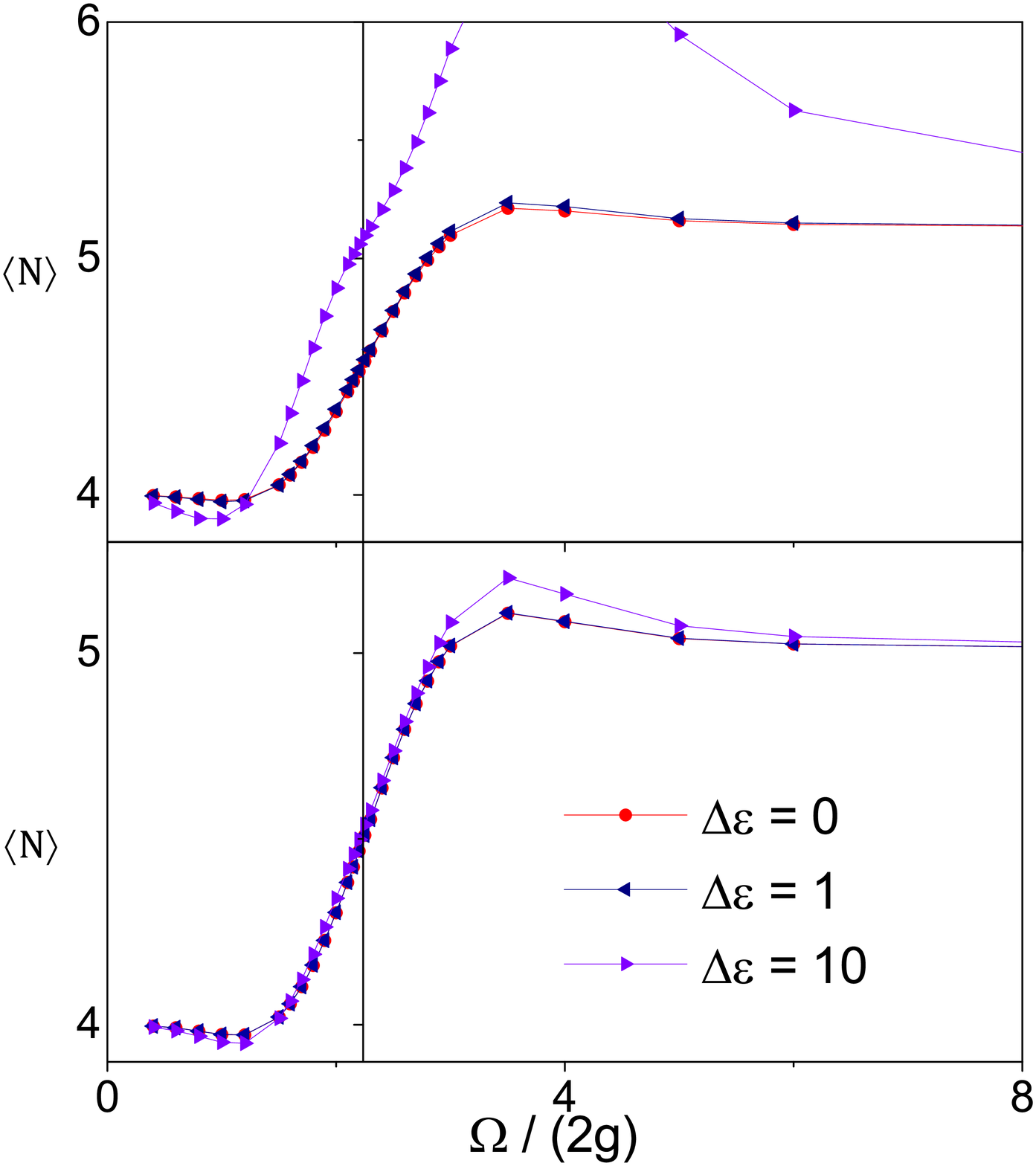}
	\caption{(color online) Results of simulation for non-Hermitian quantum walk with different system detuning $\Delta \epsilon$'s under initial $N=5$, with (upper) under qubit decay $\gamma = 4$ and (lower) $\gamma = 20$. Notice that the difference between $\Delta \epsilon = 1$ and $\Delta \epsilon = 0$ in (lower) is so tiny that one can barely distinguish them in the plot. The variable $\Omega / 2g$ captures the relative strengths of inter and intra unit hopping in the quantum walk, and $\langle N \rangle$ is the average photon number upon qubit decay. The other parameter is $g = 1$ while $\Omega$ varies.}
	\label{figEnergyGap}
\end{figure}

And Fig. \ref{figEnergyGap} shows the effect of detuning on this experiment.
Ideally, we want $\Delta \epsilon=0$, which is an exact match to the quantum walk scenario. Simulation shows that as long as the detuning of the system is kept small (Fig. \ref{figEnergyGap} shows that$\Delta \epsilon < g, \Omega$ suffices), it won't have a tangible effect on the result of this experiment.

It's worthwhile to notice that in theory, neither $\Delta \epsilon$ nor $\gamma$ should affect the experimental results. Yet in Fig. \ref{figEnergyGap}, one can observe that for a fixed $\Delta \epsilon = 10$, increasing qubit decay $\gamma$ yields better topological transition. We believe approximation (\ref{eq:approx}) plays a major role here, just like in Fig. \ref{figGamma}, as small qubit decay $\gamma$ yields wilder amplitude span to overturn approximation (\ref{eq:approx}).

\subsection{Dephasing Effects}
Dephasing characterizes one major effect of noise on circuit QED systems, making a quantum system less ``quantum'' but more ``classical''. Here, we consider dephase of the qubit due to external field fluctuation, which is one of the major sources of dephase in a circuit QED system, and can be written in Lindblad master equation (\ref{eq:master}),
where $d$ characterizes the strength of such dephase. Such dephase keeps the diagonal elements of $\rho$, but in addition to other standard evolution the non-diagonal elements of it shrinks exponentially by $e^{-d\cdot t}$.
 Here, we run a numerical simulation where qubit decay rate $\gamma$ and other parameters like $g$, $\Delta \epsilon$ are kept identical for different qubit dephase factor $d$'s.

\begin{figure}[hbtp]
	\centering
	\includegraphics[width=0.4\textwidth]{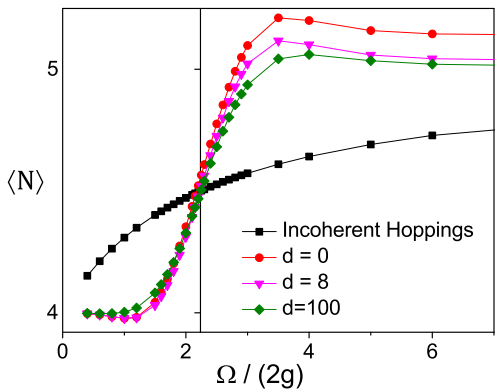}
	\caption{(color online) Results of simulation for non-Hermitian quantum walk with different qubit dephase rate $d$'s under initial $N=5$. The  variable $\Omega / 2g$ captures the relative strengths of inter and intra unit hopping in the quantum walk, and $\langle N \rangle$ is the average photon number upon qubit decay. The parameters are (same as Fig. \ref{figN100})$g = 1, \gamma = 4, \Delta \epsilon  = 10^{-4}$ while $\Omega$ varies.}
	\label{figDephase}
\end{figure}

The results are in Fig. \ref{figDephase},
which shows that the topological transition grows less distinct as the qubit dephase factor $d$ grows. Yet even when the dephase of the qubit is very large ($d=100$) compared with the other elements of the system $\Omega, \gamma, g < 10$, good topological transition can still be observed, as compared to the classical incoherent hopping.

The numerical results agree with analytical theory which we put forward earlier in this paper. This topological structure is well protected from the dephase. No matter how much dephase we have in the system, as long as such dephase comes from external field fluctuation, and can be treated as Markovian in the timescale of other operations in the system, it won't affect experimental results.

\section{2-Dimensional Quantum walk, and its realization in a circuit QED system}
\begin{figure}[hbtp]
	\centering
	\includegraphics[width=0.4\textwidth]{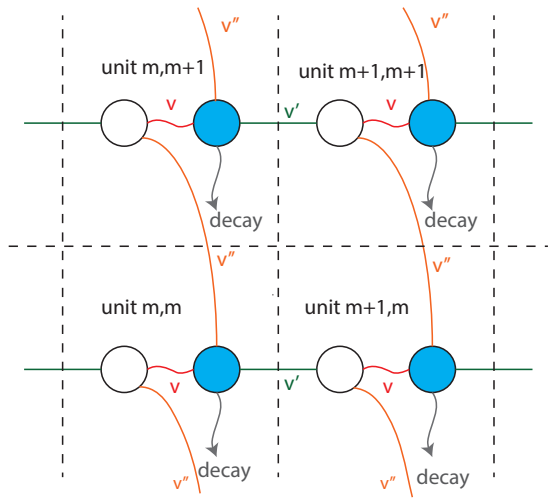}
	\caption{(color online) Two dimensional quantum walk model. The system is similar to the one in Fig. \ref{figTheory}, except with one more dimension. Three colors (red, orange and green) are used to distinguish different kinds of hopping (red for intra-site, orange and green for inter-site on two different dimensions). The dashed lines are used to indicate the boundaries of each site.}
	\label{fig2DSetup}
\end{figure}

In this section we extend the one-dimensional quantum walk to higher dimension. We first consider the 2-dimension case. Here each unit still consists of a decaying site and a non-decaying site, however, a ``walker'' could make inter-unit hopping in two dimensions, with their strengths characterized by $v'$ and $v''$ each. The strength of hopping between two sites of the same unit is still $v$. Fig. \ref{fig2DSetup} provides a sketch of the model.

In a circuit QED system, it's relatively easy to accommodate this change by adding a second resonator mode coupled to the qubit. Like before, if one uses $\ket{e/g} \otimes \ket{n_1} \otimes \ket{n_2}$ to identify the state of the system, then the Hamiltonian of the system is
\begin{eqnarray} H = & g_1 \left(\sigma^+ \tilde{a_1} + \sigma^- \tilde{a_1} ^\dagger\right) + g_2 \left(\sigma^+ \tilde{a_2} + \sigma^- \tilde{a_2} ^\dagger\right) \notag\\&+ \Omega/2(\sigma^+ + \sigma^-) + \frac{1}{2} \Delta \epsilon \sigma^z - \frac{1}{2} \frac{i\gamma}{2} \ket{e} \bra{e}, \end{eqnarray}
 where $\sigma$'s are still Pauli operators on the qubit, and $\tilde{a_{1/2}}$'s are lowering / rising operators on the first or second resonator field in a rotating frame, $\Delta \epsilon$ is (real) detuning of the system, and $\gamma$ is the decay rate of the excited qubit state.

\subsection{Analytical Theory for Higher Dimensional Quantum Walk}
The theory for higher dimensional case without dephase has been explained in \cite{RL2}, which is a simple extension of the one-dimensional case, and so is the following section considering dephase.

Suppose the dimension is $d$. As in the one-dimensional case, one can still use $\rho_{\mathbf{n,n'}}^{c,c'}$ to denote the state of the system, where $\mathbf{n}$ and $\mathbf{n'}$ are $d$-dimensional vectors and $c, c'\in \{g,e\}$.
The same Fourier Translation into momentum space yields that for subspace $\mathbf{k}$,
\begin{equation}
H_{\mathbf{k}}
=
\left(
\begin{array}{cc}
0 & \mathbf{A_k} \\
\mathbf{A_k}^* & \Delta \epsilon - i\hbar \gamma / 2 \\
\end{array}
\right)
\end{equation} with $\mathbf{A_k} = \frac{\Omega}{2} + \sum_{\alpha = 1}^d g^{(\alpha)}\sqrt{N}e^{-i k_\alpha}$. Defining $p_\mathbf{k}(t)$ in the same way as \begin{equation} p_\mathbf{k} =\rho_\mathbf{k,k}^{g,g} + \rho_\mathbf{k,k}^{e,e} \end{equation} and with $\partial_t p_\mathbf{k} = -\gamma  \psi_\mathbf{k,k}^{e,e}(t)$ and integrate by part, one can reach
\begin{equation} \langle \Delta n_\alpha \rangle =\oint \frac{d^{d-1} k}{(2\pi )^{d-1}} \left\{i\gamma \int_0^\infty dt \oint \frac{dk_\alpha}{2\pi} \partial_{k_\alpha, 1} \rho_{\mathbf{k,k}}^{e,e}\right\}, \label{BASICE2}\end{equation} while the shorthand $\partial_{k_\alpha, 1}$ means taking partial derivative only on the $\alpha$-th component of $\mathbf{k}$, and only on the first $\mathbf{k}$ in the density matrix. It's not hard to see that the expression inside the brace is exactly as equation (\ref{BASICE1}) and thus is either $0$ or $1$, no matter what dephasing factors we have.

\begin{figure}[hbtp]
	\centering
	\includegraphics[width=0.39\textwidth]{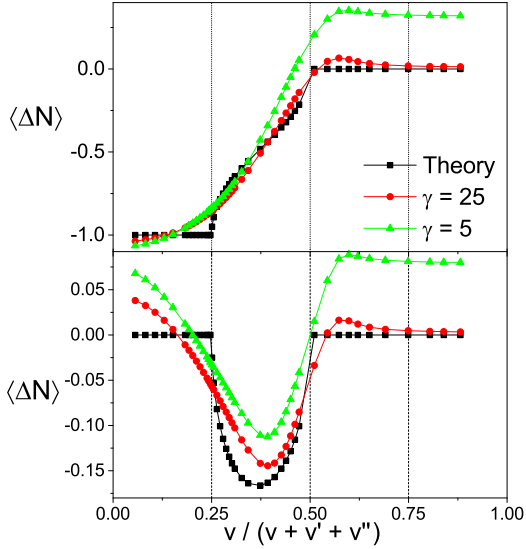}
	\caption{(color online) 2D Hoppings: Results of simulation with different qubit decay factor $\gamma$'s under initial $N = 5$, and comparison with the theoretical (ideal) case for a 2-dimensional walk in (upper) the first resonator mode, and (lower) the second resonator mode. The coupling strengths are set to $g_1 = 2, g_2 = 1$ to satisfy constrain $v'/v'' = 2$. The independent variable is calculated by $v/(v+v'+v'') := \Omega / (\Omega + 2\sqrt{N}(g_1 + g_2))$ to fit in the form of equation (\ref{eq:2DResult}). $\langle \Delta N \rangle := \langle N \rangle - N$ is the average photon number change upon qubit decay in each resonator mode. The other parameter is $\Delta \epsilon = 10^{-4}$ while $\Omega$ varies. From the fact that a similar qubit decay  ($\gamma = 5$) as 1d hoppings leads to large deviation to the ideal topological structure one can observe that a larger qubit decay is necessary for 2d experiments.}
	\label{fig2Dsim}
\end{figure}

Now, if one come back to the definition of $\mathbf{A_k} = \frac{\Omega}{2} + \sum_{\alpha = 1}^d g^{(\alpha)}\sqrt{N}e^{-i k_\alpha}$, one way to understand equation (\ref{BASICE2}) is to first fix $d-1$ angles $k_{\beta \neq \alpha}$, then see whether the integration of $k_\alpha$ from $0$ to $2\pi$ would case  the angle of $\mathbf{A_k}$ to also shift $2\pi$, and integration over those $d-1$ angles. During the integration of the other $d-1$ angles, we may observe new topological structures, like the middle area of Fig. \ref{fig2Dsim}, with some old traits still remaining, like the two sides of Fig. \ref{fig2Dsim}.

For the 2-dimensional case that is relatively easy to simulate,assuming $g_1 > g_2$, the results are
\begin{equation}\label{eq:2DResult}
    \langle \Delta n_1 \rangle=
    \begin{cases}
      -1 \\
      -1 + \frac{\theta_1}{\pi} \\
      0
    \end{cases}
    , \langle \Delta n_2 \rangle =
    \begin{cases}
    0 & v' >v + v'' \\
    -\frac{\theta_2 }{\pi} & |v-v'|\leq v'' \\
    0 & v' < v - v''
    \end{cases},
\end{equation} with $\cos \theta_1 = \frac{N(g_1^2 - g_2^2)-\Omega^2/4}{\Omega \sqrt{N} g_2}, \cos \theta_2 =  \frac{N(g_1^2 - g_2^2)+\Omega^2/4}{\Omega \sqrt{N} g_2}$, $v = \Omega / 2, v' = \sqrt{N} g_1, v'' = \sqrt{N} g_2$ in this system.

\subsection{Numerical Results}

We run numerical simulations for the 2-dimensional case in a similar manner as the 1-d case, except that the tensor product of two vectors is stored. We truncated photon number at $\text{MAXN} = 20$ with hindsight knowledge from the 1-dimensional simulations.  From Figure \ref{fig2Dsim} one can see that the results of the simulation preserve the basic properties of a 2-d transition. It also shows that in the 2-dimensional case, a more stringent requirement is put onto the system as a much bigger decay factor $\gamma$ is necessary to preserve the topological nature of the system compared to the one-dimensional case.

\begin{figure}[hbtp]
	\centering
	\includegraphics[width=0.39\textwidth]{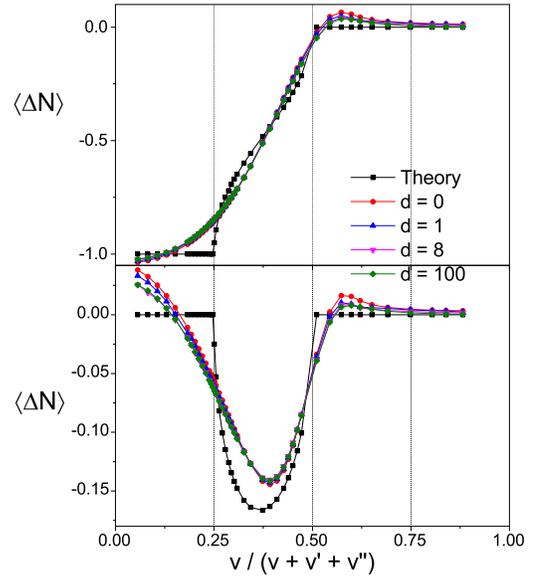}
	\caption{(color online) 2D Hoppings: Results of simulation with different dephase factor $d$'s for the qubit under initial $N=5$, and comparison with the theoretical (ideal) case for a 2-dimensional walk in (upper) the first resonator mode, and (lower) the second resonator mode. The coupling strengths are set to $g_1 = 2, g_2 = 1$ to satisfy constrain $v'/v'' = 2$. The independent variable is calculated by $v/(v+v'+v'') := \Omega / (\Omega + 2\sqrt{N}(g_1 + g_2))$ to fit in the form of equation (\ref{eq:2DResult}). $\langle \Delta N \rangle := \langle N \rangle - N$ is the average photon number change upon qubit decay in each resonator mode. The other parameters are $\gamma = 25, \Delta \epsilon  = 10^{-4}$ while $\Omega$ varies. From the fact that the curves for different dephase factor $d$'s almost overlap one can see that qubit dephase factor doesn't affect the simulated result.}
	\label{fig2DDephase}
\end{figure}

We have also numerically simulated the effect of qubit dephase on the final result. Using a similar method with section IV.B, we simulated equation (\ref{eq:master}) exactly and displayed the results in Fig. \ref{fig2DDephase}, which shows that different qubit dephase factor $d$'s doesn't affect our result.

Due to the fact that analytical theory for different dimensions follow similar integration over contours, we expect that in higher dimensional case, the topological effect would still be protected from qubit dephase.

\section{Conclusion}

In summary, we considered the quantum walk on an one-dimensional bipartite indexed lattice, where
each unit has one decay site and one non-decay site. We have proved the topological transition of average decay site in this non-Hermitian system is not affected by either the decay or the dephasing
between sites within units.

We proposed a circuit QED model where a qubit is coupled to one resonator mode. We have numerically shown that such topological transition in a non-Hermitian quantum walk can be realized in this system.
We numerically proved that neither large decay nor dephase of the qubit would affact the topogical transition. In fact, we found that the qubit decay should be larger than $g$, in order to maintain the topological transition.
We extended this circuit QED implementation to higher dimension quantum bipartite walks with novel features that arise with higher dimensionality.

We would like to thank L. Sun for helpful discussions.
This work is funded by the NBRPC (973 Program) 2011CBA00300 (2011CBA00301), NNSFC NO.61435007, NO. 11574353,
NO. 11274351 and NO.11474177.

\appendix
\section{ Detailed Deduction From Quantum Jump to Non-Hermitian Hamiltonian}
As with our main text, we assume that qubit decay and qubit dephase are the only sources of decoherence in the system. While qubit dephase is undetectable from photon detectors, qubit decay can be detected with leaking photon. For our specific system, we followed the procedure adopted in quantum-jump simulations \cite{Plenio98} \cite{R1} \cite{R2} for our system.

Suppose the system starts in state $\rho$. In order to determine the evolution of the system, we take these steps:

1. Determine the current probability of a qubit decay over an infinitesimally small time $\Delta t$.
$$ \Delta P = \gamma \Delta t Tr\left[\sigma^- \rho \sigma^+\right].$$

2. Obtain a random number $r$ between zero and one, compare with $\Delta P$, and decide on emission as follows:

3. If $r<\Delta P$ there's an emission, so that the system jumps to the renormalized form
\begin{equation}\rho \rightarrow \frac{\sigma^- \rho \sigma^+}{Tr\left[\sigma^- \rho \sigma^+\right]}.\end{equation}

4. If $r>\Delta P$, no emission takes place, so the system evolves under original Hamiltonian, without the decay term
\begin{equation}
\rho \rightarrow
\rho - i [H_{org}, \rho]\Delta t + d \Delta t \left(\sigma^z \rho \sigma^z - \rho\right).\label{eq:main}\end{equation}

5. Repeat to obtain an individual trajectory, or history.

6. Average observables over many such trajectories.

We know that the system splits into two alternatives in a time $\Delta t$, that is, into $\rho_{decay}$ with probability $\Delta P$ and into $\rho_{no-decay}$ with probability $1-\Delta P$. To show that the non-Hermitian Hamiltonian has the same effect as the system as master equation, we consider the evolution of the total density matrix, as $\rho$ evolves into
\begin{align}
& \Delta P\rho_{decay} + (1-\Delta P) \rho_{no-decay}\\
= & \gamma \Delta t Tr\left[\sigma^- \rho \sigma^+\right] \cdot \frac{\sigma^- \rho \sigma^+}{Tr\left[\sigma^- \rho \sigma^+\right]}+ (1-\gamma \Delta t Tr\left[\sigma^- \rho \sigma^+\right]) \notag  \\ & \cdot \left(\rho - i [H_{org}, \rho]\Delta t + d \Delta t \left(\sigma^z \rho \sigma^z - \rho\right)\right)\\
 \approx &  \rho - i [H_{org},\rho] \Delta t + d \Delta t \left(\sigma^z \rho\sigma^z - \rho\right)  \notag  \\ &+ \gamma \Delta t \left(\sigma^- \rho \sigma^+  - \rho Tr\left[\sigma^- \rho \sigma^+\right] \right)
\end{align}

Since we have
\begin{equation}  \rho Tr\left[\sigma^- \rho \sigma^+\right] =\sigma^+ \sigma^- \rho = \rho  \sigma^+ \sigma^-,\label{eq:ref}\end{equation}
one can see that this is exactly how the system would evolve under master equation (\ref{eq:master_org}).

On the other hand, in Eq. \ref{eq:main}, one can see that with non-Hermitian Hamiltonian
\begin{equation} H = H_{\text{org}} - \frac{i\gamma}{2} \sigma^+ \sigma^- \end{equation} and the relation in Eq. \ref{eq:ref} we have
\begin{align}
& (1-\gamma \Delta t Tr\left[\sigma^- \rho \sigma^+\right])\left(\rho - i [H_{org}, \rho]\Delta t + d \Delta t \left(\sigma^z \rho \sigma^z - \rho\right)\right)\notag \\ \approx & \rho - i [H, \rho]\Delta t + d \Delta t \left(\sigma^z \rho \sigma^z - \rho\right),
\end{align}
where the approximation holds in the first order of $\Delta t$. The left hand side is just Eq. \ref{eq:main}, multiplies the probability that the qubit doesn't decay; the right hand side is evolution under the non-Hermitian Hamiltonian. This shows the non-Hermitian Hamiltonian $H$ characterizes the evolution of the system given that the qubit doesn't decay, and the probability of that. So, in our paper we used
\begin{equation} \dot{\rho}(t) = -i [H,\rho(t)] +d \left(\sigma^z \rho(t) \sigma^z - \rho(t)\right) \end{equation} to characterize the evolution of the system.




\begin{thebibliography}{99}
\bibitem{FOOTNOTE3} A simple explanation based on the first decay site of the walker is as follows: with probability $\frac{v}{v'+v}$ the first decay site of the walker is on the same unit, with probability $\frac{v'}{v'+v}$ the first decay site of the walker is on the previous unit. Once a walker is on a decay site, the average decay site from then on is just that unit based on the symmetry of the system. So, the average displacement would be $-\frac{v'}{v'+v}$.
\bibitem{FOOTNOTE1} Although with hindsight we only need the dimension of Hilbert space for the resonator to be $2N$, which means  $\text{MAXN} \sim 4 N$ is simply enough. Choosing such a large number is to make sure that we don't need to change the configuration of the system for different initial Fock state numbers like $N=1$ and $N = 100$.
\bibitem{FOOTNOTE2} Due to the existence of ``dark state'' for finite systems, that amplitude won't go to exactly zero.
\end{thebibliography}
\end{document}